\documentclass{article}

\usepackage{arxiv}

\usepackage[utf8]{inputenc} % allow utf-8 input
\usepackage[T1]{fontenc}    % use 8-bit T1 fonts
\usepackage{hyperref}       % hyperlinks
\usepackage{url}            % simple URL typesetting
\usepackage{booktabs}       % professional-quality tables
\usepackage{amsfonts}       % blackboard math symbols
\usepackage{nicefrac}       % compact symbols for 1/2, etc.
\usepackage{microtype}      % microtypography
\usepackage{lipsum}		    % Can be removed after putting your text content
\usepackage{graphicx}
\usepackage[square,sort,comma,numbers]{natbib}
\usepackage{doi}
\usepackage{xcolor}
\newcommand{\be}{\begin{equation}}
\newcommand{\ee}{\end{equation}}
\newcommand{\bc}{\begin{center}}
\newcommand{\ec}{\end{center}}
\newcommand{\bi}{\begin{itemize}}
\newcommand{\ei}{\end{itemize}}
\newcommand{\ba}{\begin{eqnarray}}
\newcommand{\ea}{\end{eqnarray}}

\newcommand{\ignore}[1]{}

\title{Untangling the brain web: from the early days of complex functional networks to  the non-linear dynamical directed functional connectivity measures}

\author{ \href{https://orcid.org/0000-0002-1038-3637}{\includegraphics[scale=0.06]{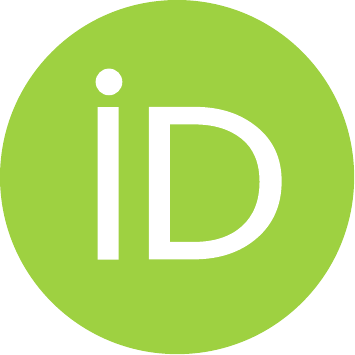}\hspace{1mm}Dante R. Chialvo} \\
	Center for Complex Systems \& Brain Sciences \\(CEMSC$^3$), \\
	Escuela de Ciencia y Tecnolog\'ia,  \\
	Universidad Nacional de San Mart\'{i}n, \\ 
	Buenos Aires, Argentina \\
	\\
	Consejo Nacional de Investigaciones Cient\'{i}ficas\\ 
	y Tecnol\'{o}gicas (CONICET),\\
	Buenos Aires, Argentina.\\
	\texttt{dchialvo@gmail.com} \\
	\And
	\href{https://orcid.org/0000-0003-0644-6022}{\includegraphics[scale=0.06]{Texfiles/orcid.pdf}\hspace{1mm}Ignacio Cifre} \\
	Facultat de Psicologia,\\ Ci\`encies de l'educaci\'o  i de l'Esport,\\
	Blanquerna, Universitat Ramon Llull, \\
	Barcelona, Spain.\\
	\\
	Center for Complex Systems \& Brain Sciences \\(CEMSC$^3$), \\
	Escuela de Ciencia y Tecnolog\'ia,  \\
	Universidad Nacional de San Mart\'{i}n, \\ 
	Buenos Aires, Argentina \\
	\texttt{ignaciocl@blanquerna.url.edu} \\
	
	\And
	\href{https://orcid.org/0000-0002-7281-1852}{\includegraphics[scale=0.06]{Texfiles/orcid.pdf}\hspace{1mm}Jeremi K. Ochab} \\
	Institute of Theoretical Physics\\
	and\\
	Mark Kac Center for Complex Systems Research,\\
	Jagiellonian University, Poland\\
	\texttt{jeremi.ochab@uj.edu.pl} \\
}

\renewcommand{\shorttitle}{nonlinear
brain functional connectivity}

\hypersetup{
pdftitle={brain functional connectivity},
pdfsubject={q-bio.NC, q-bio.QM},
pdfauthor={Dante R. Chialvo, Ignacio Cifre},
pdfkeywords={First keyword, Second keyword, More},
}

\begin{document}
\maketitle

\begin{abstract}
Already two decades passed since the first applications of graph theory to brain neuroimaging. Since that early description, the characterization of the brain as a very large interacting complex network has evolved in several directions. In this brief review we discuss our contributions to this topic and discuss some perspective for future work. 
\end{abstract}


\section{Introduction}

According to mainstream ideas, the brain can be studied as a very large interacting complex network in which each region of interest (or even a voxel) is considered a node. In such a view, brain functioning results from the interactions of the nodes through flexible links connecting them transiently. Despite the simplicity behind this enunciation, its realization in practical terms took two decades, and it is still evolving. In this note\footnote{A contribution to celebrate the Jubilee of Dr. Tadeusz Marek, professor of psychology at the Jagiellonian University, Chairman and Cofounder of the Neurobiology Department / NeuroImaging Group of Malopolska Centre of Biotechnology and the Department of Cognitive Neuroscience and Neuroergonomics, Institute of Applied Psychology in Krakow, Poland.} we review the evolution of these ideas, highly biased toward our own contributions and viewpoints. The paper is organized as follows: The following section describes the efforts coinciding with the upheaval of enthusiasm created by the application of graph theory ideas at the beginning of the millennium.  Section 3 will dwell into our efforts to drift away from the first network definition by introducing the concept of Resting BOLD Evoked Triggered Activity (rBeta)\cite{rbeta}. Section 4 will describe how to get access to fast dynamical changes in the functional connectivity by transforming the BOLD signal into a point process \cite{Tagliazucchi2012, Tagliazucchi2016,Cifre2020}. In Section 5 we revisit the rBeta technique to highlight new properties able to describe dynamical measures of functional connectivity.  The paper closes with a discussion of a set of methods developed by others and how they relate with each other and with our own work.  

\section{Graph theory upheaval: The early days of functional brain networks}

With the turn of the millennium, the application of graph theory took by storm almost any scientific discipline \cite{albert2002,strogatz2001,watts1998}, Neuroscience was not an exception. The first applications of graph theory to the brain were dedicated to describing the brain structural connectivity, a topic championed by Sporns and colleagues, as documented in great detail in Ref.~\cite{SpornsBook}. The initial work included the statistical analyses of small data sets of \textit{C. Elegans} \cite{watts1998}, and two neuroanatomical brain databases \cite{sporns2000,hilgetag2000}, the macaque visual cortex \cite{felleman1991} and the cat cortex \cite{scannel1999}. 
Concerning the application of graph theory to brain activity (i.e., functional), the work of Stam \& de Bruin \cite{stam} was the first to derive graphs from electroencephalogram time-series. The first  very large functional brain network was described by Eguiluz and colleagues, sometime during 2003 and published in \cite{Eguiluz2005} studying time-series of Blood Oxygenated Level Dependent (BOLD) signals measured with functional magnetic resonance imaging (fMRI) data.  Such seminal work introduced the term ``complex functional brain network''  to name the network defined by the strongest correlated brain voxels.

The cartoon in Figure 1 shows how the functional brain networks were constructed  in Ref. \cite{Eguiluz2005}. In these experiments, at each time step
(typically 400 samples spaced 2.5 sec.), the magnetic resonance brain activity
was measured in $36\times64\times64$ brain sites (so-called ``voxels'' of dimension $3\times 3.475\times 3.475~$mm$^3$). 
The activity of a voxel $x$ at time $t$ was denoted as $V(x,t)$. Two voxels were considered {\sl functionally connected} if their temporal correlation exceeded a positive pre-determined value $r_c$, regardless of their {\sl anatomical connectivity} \cite{dodel2002}.
Networks can also be defined using negative correlations, but for simplicity we restrict ourselves here to the positive case only. Specifically, the linear correlation coefficient was computed between any pair of voxels, $x_1$ and
$x_2$, as: \be r(x_1, x_2)= \frac{\langle V(x_1,t) V(x_2,t)\rangle
- \langle V(x_1,t)\rangle \langle V(x_2,t) \rangle} {\sigma
(V(x_1))\sigma (V(x_2))} ~, \ee where $\sigma^2(V(x)) = \langle
V(x,t)^2\rangle - \langle V(x,t)\rangle^2$, and $\langle \cdot
\rangle$ represents temporal averages.

%%%%%%%%%%%%%%%%%%%%%%%%%%%%%%%%%%%%%%%%%%%%%%%%%%%%%Fig.1
\begin{figure}[ht!]
\centering
\includegraphics [width = .7 \linewidth] {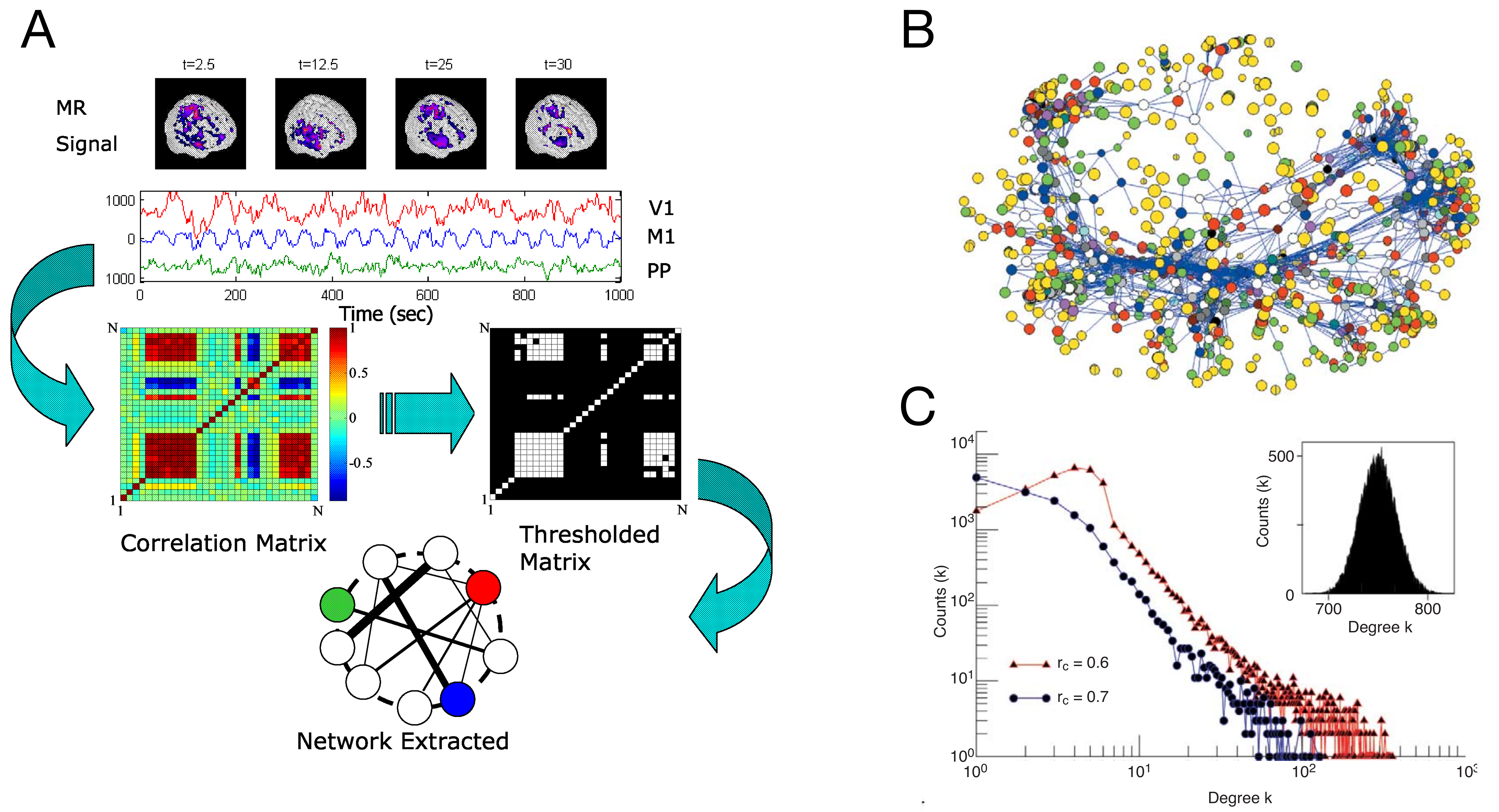} 
\caption{\footnotesize{Methodology introduced in 2003 by Eguiluz et al.\cite{Eguiluz2005} and now widely used to extract functional brain networks from BOLD signals. Panel A: The correlation matrix from all the signals is computed according to Eq. 1 and then thresholded to define a voxel-wise network among the highest correlated nodes. The top four images represent snapshots (at times t=2.5, 12.5, 25, and 30 secs.) of brain activity, and the three traces correspond to the BOLD time-series of selected voxels from visual (V1), motor (M1) and posterior-parietal
(PP) cortices. Panel B: A typical functional brain network extracted from human fMRI data. Nodes are colored according to their degree (yellow=1, green=2, red=3, blue=4, other colors > 4). (d) Degree distribution for the two correlation thresholds denoted in the legend. The inset depicts the degree distribution expected for an equivalent random network. Data from \cite{Eguiluz2005}. }}
\end{figure}
%%%%%%%%%%%%%%%%%%%%%%%%%%%%%%%%%%%%%%%%%%%%%% 
 
The functional brain networks exhibited scale-free features, both at the level of the node degree distribution as well as in the decay of the linear correlation as a function of Euclidean distance, features that were widely confirmed, starting with the work of  van~den Heuvel \cite{VandenHeuvel2008}. The presence of these features anticipated the existence of hubs (i.e., very well-connected nodes)  and rich club organizations, topics that have been profusely studied since then. Figure 2 shows two excerpts selected from the multitude of reports confirming the scale freeness of correlations uncovered in the original work of Eguiluz et al. Panels A and B  belong to the work of van~den Heuvel \cite{VandenHeuvel2008} showing a scale-free degree distribution (in  Panel A) and the regions exhibiting the highest connectivity degree, i.e. the voxels that showed the largest number of functional connections (Panel B). These hubs included the right and left thalamus, bilateral superior temporal lobe (BA 22/40/42), bilateral anterior cingulate cortex (BA 24) and bilateral posterior cingulate cortex/(pre)cuneus (BA 30/31/18). Later on, Tomasi and Volkow \cite{Tomasi} replicated these findings over a large database. As shown in Fig. 2 C, on average, similar hubs were systematically identified across subjects with high test-retest reproducibility.

 %%%%%%%%%%%%%%%%%%%%%%%%%%%%%%%%%%%%%%%%%%%%%%%%%%%%%Fig.1
\begin{figure}[ht!]
\centering
\includegraphics [width = .6 \linewidth] {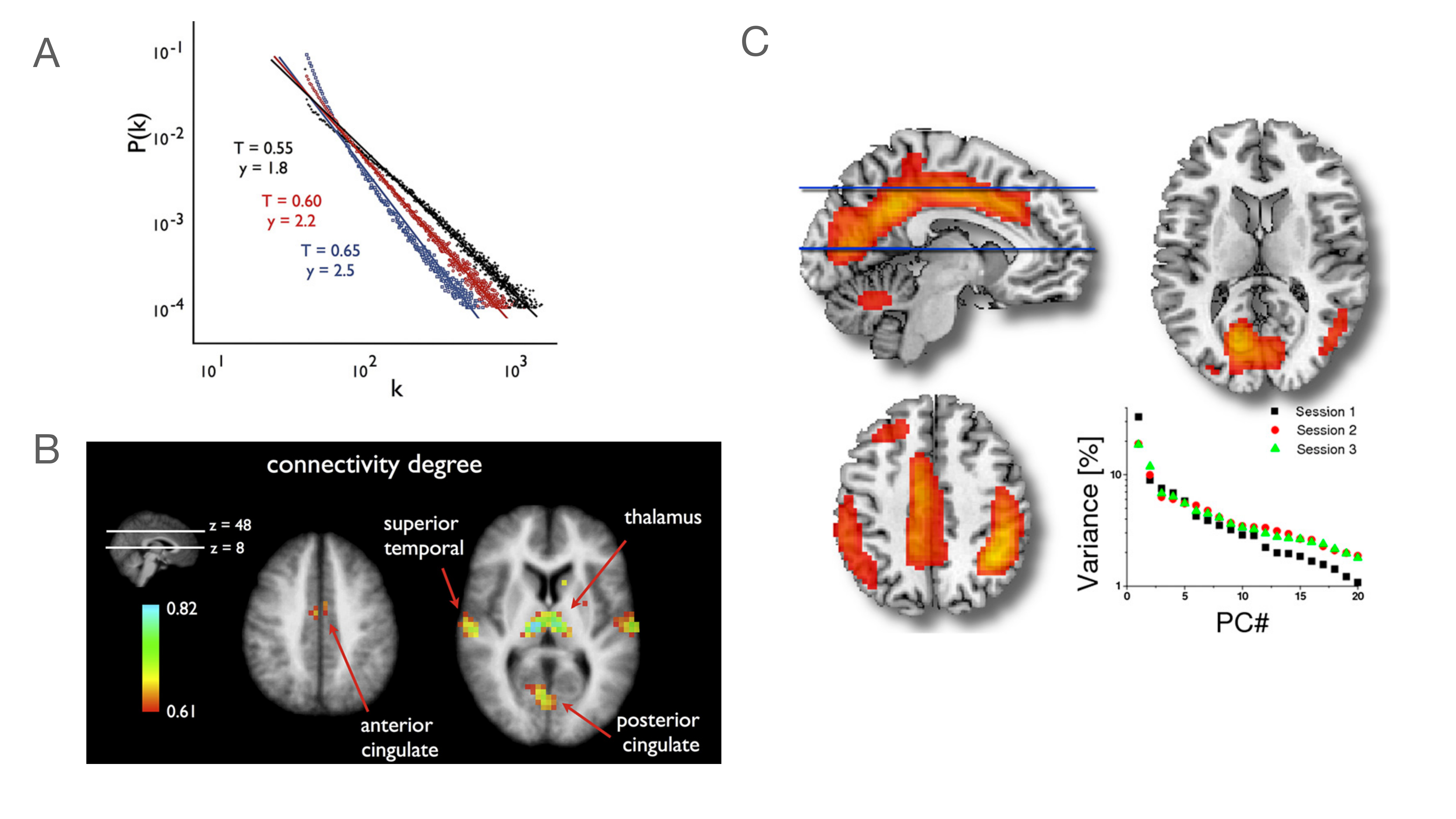} 
\caption{\footnotesize{Panels A and B show, respectively,  the node degree distribution across the brain for three values of the threshold used to define the presence of an edge and the brain locations of the hubs (from \cite{VandenHeuvel2008}).  Panel C: Tomasi \& Volkow's \cite{Tomasi} mapping of the high degree regions. Average spatial distribution of the first principal component (PC 1) demonstrating the brain regions with the highest degree (i.e., hubs). The bottom-left plot demonstrates the high test-retest reproducibility (\% of variance) as a function of the principal components for each of the three sessions. }}
\end{figure}
%%%%%%%%%%%%%%%%%%%%%%%%%%%%%%%%%%%%%%%%%%%%%% 

\subsection{From $\sim$ 40000 voxels to $\sim$ 100 ROIs: The illusion of a small network?}

Typically, the voxel-wise complex functional brain networks introduced by Eguiluz et al. involved the calculation of several thousand correlation pairs, resulting in adjacency matrices of up to $\sim 10^8$ edges. Perhaps due to these extensive computations,  coarse-grained versions were introduced rapidly, in which only a very small fraction ($\sim$ hundreds) of regions of interest (ROI) were considered as the network's nodes. An example is the AAL parcelation \cite{AAL} which comprises 90-116 cortical/subcortical regions.
In a few years, scores of papers used these ROI-based networks to compare control and patient groups in the hope to provide a straightforward analytical tool. Overall, the results of these efforts were mixed, in many cases because such coarse-grain ignores, by definition, the correlation dynamics within each ROI. Indeed, in such an approach, all nodes are treated equally, despite the fact that some ROIs may represent very large cortical regions and others very small ones. This may explain the relatively low success of the AAL parcelation-based networks in discriminating between healthy and pathological conditions.

\subsection{The unsolved issue of selecting a correlation threshold}

As explained above, one defines functional networks by the nodes whose activity is correlated beyond a given threshold.
Such a definition brings several advantages: binary adjacency matrices take smaller storage memory, especially when they cross a certain threshold; it increases signal-to-noise ratio by eliminating the small correlations for which reliable inference is hard; the binarization disentangles topology from the connection weights.
Last but not least, it simply brings into play the whole toolbox of complex networks analysis with its numerous metrics, null models, and analytical results.

\begin{figure}[ht!]
\centering
\includegraphics [width = .6 \linewidth] {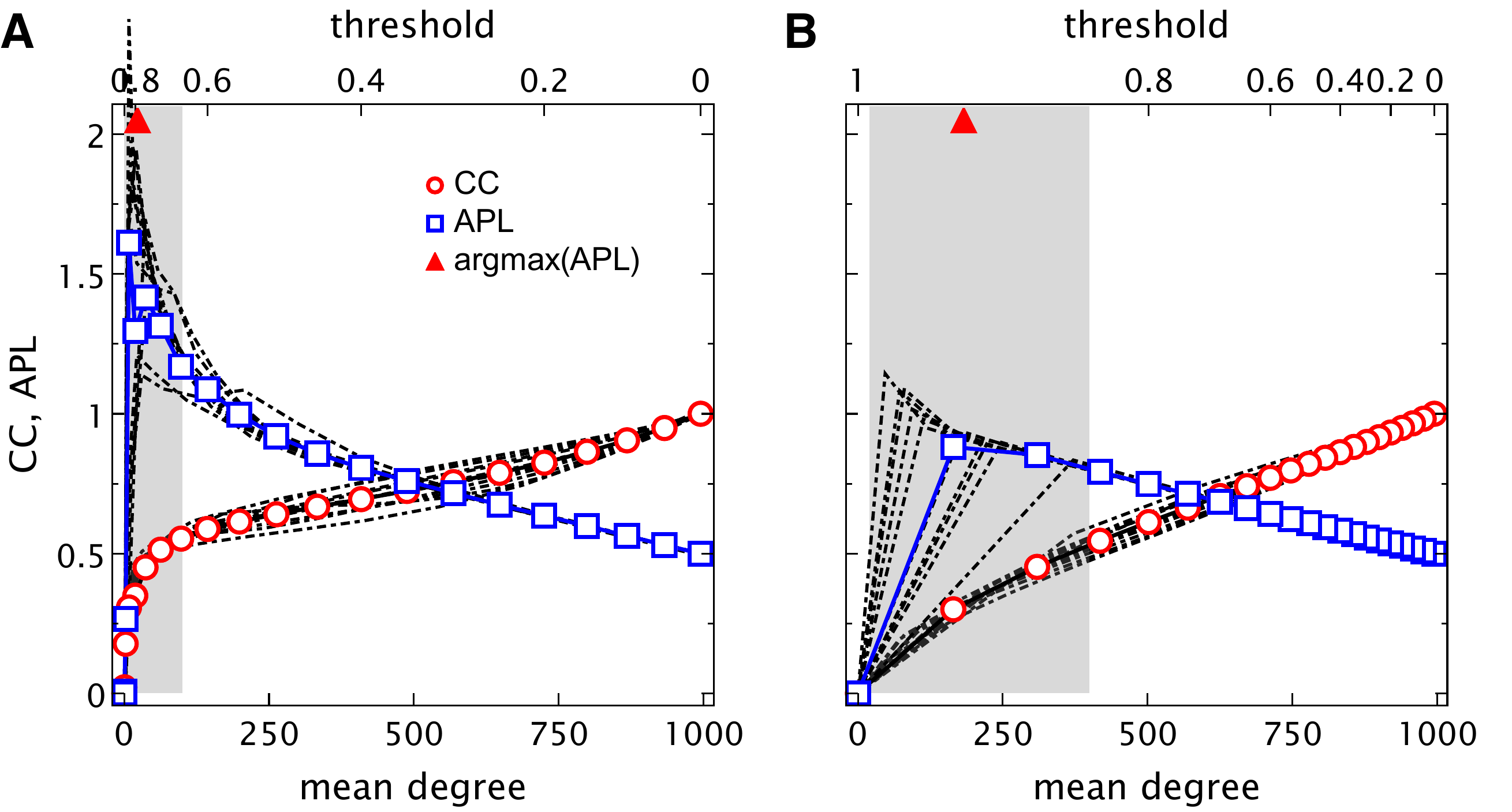}
\caption{\footnotesize{The average path length (APL; in blue) of the largest connected component and the average local clustering coefficient (CC; in red) of functional networks for Pearson correlations (Panel A), and rBeta correlations (Panel B). For a large correlation threshold (upper axis), resulting in a low density of links (lower axis), the largest component disintegrates, manifesting as a peak in APL. The shaded area is the `danger zone', where the choice of the threshold heavily influences the later outcomes. The fMRI data were 998 ROIs and 224 time points averaged over 10 subjects in resting state (the dot-dashed curves correspond to individual subjects).
}}
\label{Threshold}
\end{figure}

While selecting the appropriate value for the threshold is a critical processing step, it is not straightforward as each of the above advantages comes with a caveat.
The issue of determining the optimal threshold---and, in fact, defining what `optimal' means---has been largely ignored in the literature and has remained unsolved (for a brief overview, see chapters 3.2.1 and 11.1 in \cite{fornito2016}).
To demonstrate the relevance of this problem, we present Fig.\ref{Threshold}, with the two quantities most commonly used to characterize networks: clustering coefficient and average path length \cite{VandenHeuvel2008}.
The figure shows the effect of lowering the threshold of Pearson correlations and the rBeta correlations that we introduce in the next section.
The visible peak in the average path length (in the largest connected component; it would become infinite if computed for the disconnected graph) marks the phase transition akin to percolation transition, where a giant connected component (GCC) appears (see \cite{Bollobas2008} for a synthetic summary of known results).
That peak is where the network's characteristics are possibly the most interesting and where the signal-to-noise ratio is at its largest.
At the same time, one can expect that the sizes and other quantities describing GCC would exhibit the largest variance near the transition, making it a dangerous zone for placing the threshold.

 %%%%%%%%%%%%%%%%%%%%%%%%%%%%%%%%%%%%%%%%%%%%%%%%%%%%%Fig.3
%%%%%%%%%%%%%%%%%%%%%%%%%%%%%%%%%%%%%%%%%%%%%% 

Other caveats result from the assumptions that lie behind the idea of selecting a single optimal threshold. 
Let us spell out some of these assumptions: first, the functional network characteristics ought to be homogeneous across population and intraindividually across time.
Naturally, that assumption is not valid for BOLD correlations, as exemplified by the dot-dashed curves in Fig.~\ref{Threshold} representing ten individuals.
This fact leads to the need for normalizing the correlations on the person-by-person basis or selecting individual thresholds in order to be able to make any group comparisons, which are the basic experimental designs in medicine, psychology, and other disciplines utilizing fMRI.

Second, the correlation threshold indicative of a relevant functional association should be independent of the particular pair of ROIs.
This point is much subtler: for linear correlations, it would be true if the BOLD time series registered in all the ROIs had the same underlying generative process and consequently the same autocorrelation characteristics.
We know this not to be true in general since the autocorrelations happen to take a range of possible values and influence the value of cross-correlations \cite{Ochab2019}. Whether this leads to more robust or to spurious correlations remains debatable.

The third assumption needed for a single threshold to be feasible is the requirement that the network metrics used in any further analysis do not change, given the threshold.
This, again, becomes problematic in connection with the inter- and intra-individual variability, which for a naive threshold of correlation values would result in networks of different link densities.
Different network metrics (average path length and correlation coefficients among them), in turn, are known to depend on network size or link density in network null models \cite{vanWijk2010}.
Hence, a threshold expressed in link density (or mean degree) might provide a better basis for comparisons. That is why we plot it on the x-axis in Fig.~\ref{Threshold}.
Redefining the threshold in that way, however, comes at the cost of enforcing a variable number of false positive or false negative correlations \cite{Zalesky2016}.

The list of possible issues and false assumptions is definitely not exhaustive. 
% - assumptions: independent Gaussian noise 
There also has not appeared a universally acknowledged solution, but rather a list of approaches---each with its pros and cons (see, e.g., discussions in \cite{fornito2016,VandenHeuvel2017}).
For instance, one might base the threshold choice on significance testing for the strongest correlations.
Then, instead of one global threshold, local thresholding schemes have been designed; among them are methods based on adding links to the minimum spanning tree, thus avoiding network fragmentation. 
% - methods to choose the threshold: cross-validated, based on optimizing some network features,..., selection of elements consistent within a group \ref{Ginestet2011}
Next, it is possible to integrate the information across a range of threshold values into a single, more robust measure.
Another idea would be to either find network metrics invariant to the threshold changes or to normalize them against some null models before further analysis. 
% - which network characteristics change and how: density, clustering, path length \cite{VandenHeuvel2008}, modularity? ...
Finally, one can avoid thresholding at all, given graph metrics suited for fully weighted networks---which, then, might depend on quantities such as network density in even more complex and less controllable ways.

While we lack the solution, or perhaps a well-defined formulation of the problem, one step to at least partially safeguard any functional network analysis should happen even before constructing the network: finding a better estimator of functional correlations or denoising them. Promising inference methods but relatively understudied in this context come from random matrix theory \cite{Burda2013,Almog2019}, while one of the alternative ways of defining correlations, better grounded in fMRI physiology than linear correlations, is the topic of the following chapters.

\section{Large events are more important: the rBeta method}

The so-called hemodynamic response function (HRF) is the stereotypical BOLD pattern \cite{friston1995,Aguirre1998} recorded after any given neuronal activation. Such patterns, resembling a gamma function, were primarily studied in response to external perturbations.  

  %%%%%%%%%%%%%%%%%%%%%%%%%%%%%%%%%%%%%%%%%%%%%%%%%%%%%Fig.1
\begin{figure}[ht!]
\centering
\includegraphics [width = .5 \linewidth] {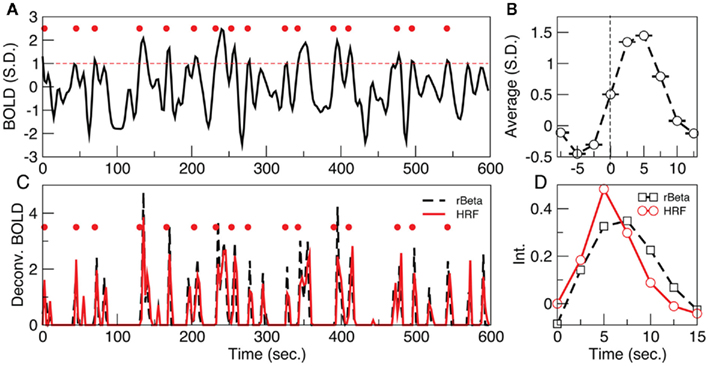} 
\caption{\footnotesize  A point process and a stereotypical average pattern (rBeta) can be extracted from the typical BOLD signal response to intrinsic neuronal activations (A) Example of a point process (filled circles) extracted from the normalized BOLD signal. Each point corresponds to a threshold (dashed line at 1 SD) crossing from below. (B) Average rBeta extracted from the BOLD signal triggered at each threshold crossing. (C) The peaks of the de-convolved BOLD signal, using either the hemodynamic response function (HRF) or the rBeta function \cite{Tagliazucchi2010} depicted in (D), coincide on a great majority with the timing of the points shown in (A). Figure reproduced from \cite{Tagliazucchi2012}}
\end{figure}

Considering that brain activity never stops, we suggested that it may be possible to define a similar HRF in response to the spontaneous ongoing neural activity \cite{Tagliazucchi2010}. Heuristically, we extracted the largest amplitude deflections of the BOLD signal by averaging epochs of the signal triggered by the threshold crossings.  Figure 3 shows an example of how the large-amplitude events are defined and how they relate with the HRF. Panel A depicts the typical BOLD time series of a  single voxel from a subject during resting state. The observation made initially in Ref. \cite{Tagliazucchi2010} is that a definitive pattern (see Fig. 3B) can be extracted from the BOLD samples subsequent to the upward crossings of a suitable threshold. Since these BOLD-evoked patterns are triggered by the signal fluctuations (i.e., there is no explicit stimulus) we call it ``resting BOLD evoked triggered activity'' or rBeta. In \cite{Tagliazucchi2012} it was demonstrated that the rBeta extracted in this manner strongly resembles the canonical grey matter HRF (see Panel D), which results in very similar de-convolved activity when using either one, as shown in Fig.3C.  The examples in Figure 4 show the temporal relationship between the source and target events, which are eventually used in this approach to compute the functional connectivity at resting state between any given pair of ROIs.

 %%%FIGURE 1 - Event selection %%%%%%%%%%%%%%%%%%%%%%%%%%%%% 
%%%%%%%%%%%%%%%%%%%%%%%%%%%%%%%%%%%%%%%%%%%%%%%%%%%%%%%%%%%%
\begin{figure}[htb]
\centering
\includegraphics[width=0.45\textwidth]{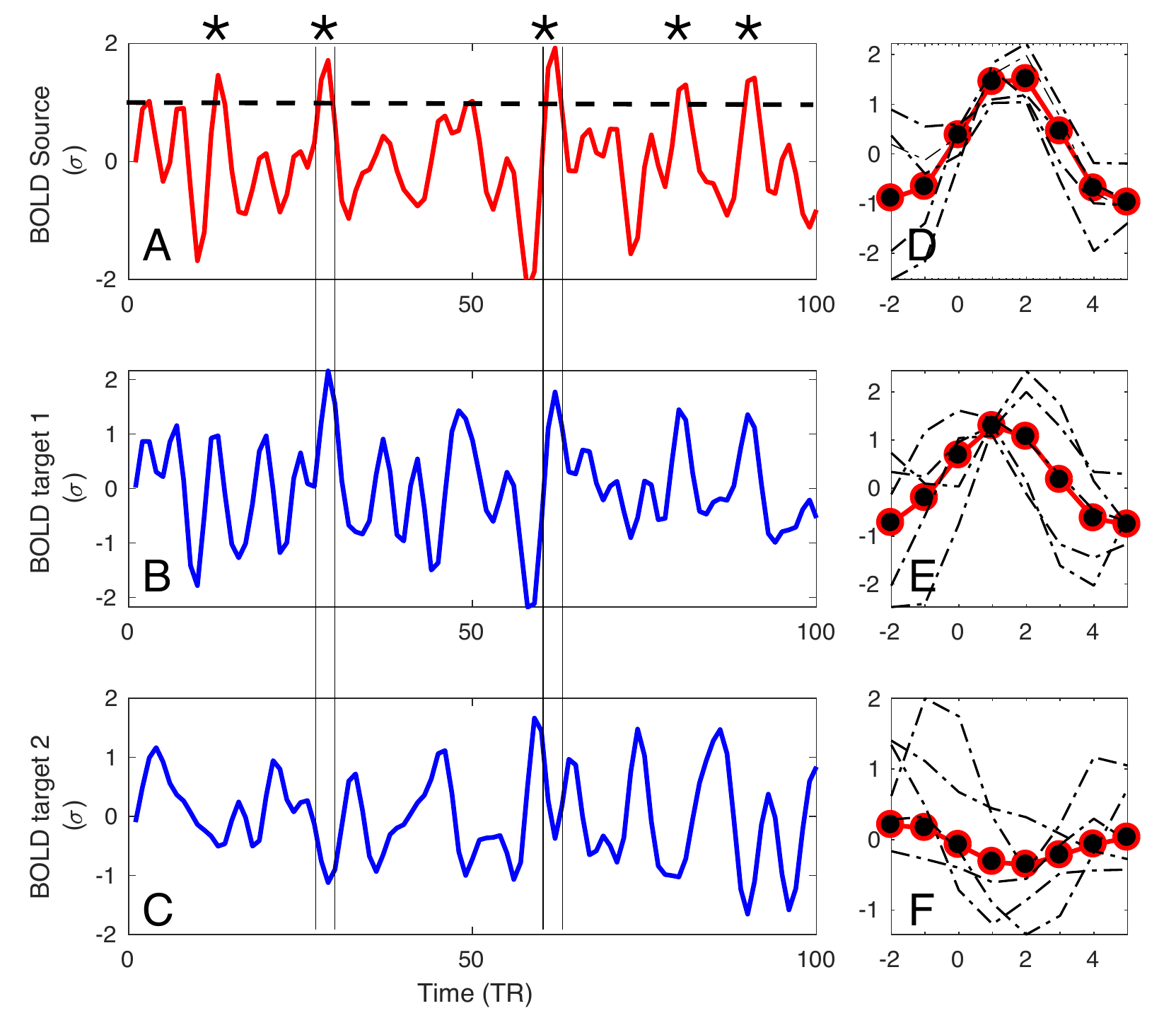}
\caption{Definition of the large amplitude BOLD events: For each source region (panel A), the BOLD triggered (or also called ``source'') events, denoted with asterisks, are defined at the times at which the BOLD signal crosses an arbitrary threshold (here set to $1 \sigma$, denoted by the dashed line in panel A). For each source event, a target event can be extracted coinciding with the times of the source from the BOLD signals of the other regions of interest (as the two examples in panels B and C denoted by vertical lines). Subsequently, both the source and target events can be averaged (Panels E-F), and used to further compute correlations, delays, and directionality. Figure from \cite{Cifre2021}}
\label{SelectEvents}
\end{figure}
%%%%%%%%%%%%%%%%%%%%%%%%%%%%%%%%%%%%%%%%%%%%%%%%%%%%%%%%%%%%

%%%%%%%%%%%%%%%%%%%%%%%%%%%%%%%%%%%%%%%%%%%%%%%%%%%%%%%%%%%%%%%%%%%%%%%%%%
\begin{figure}[ht!]
\centering
\includegraphics [width = .65 \linewidth] {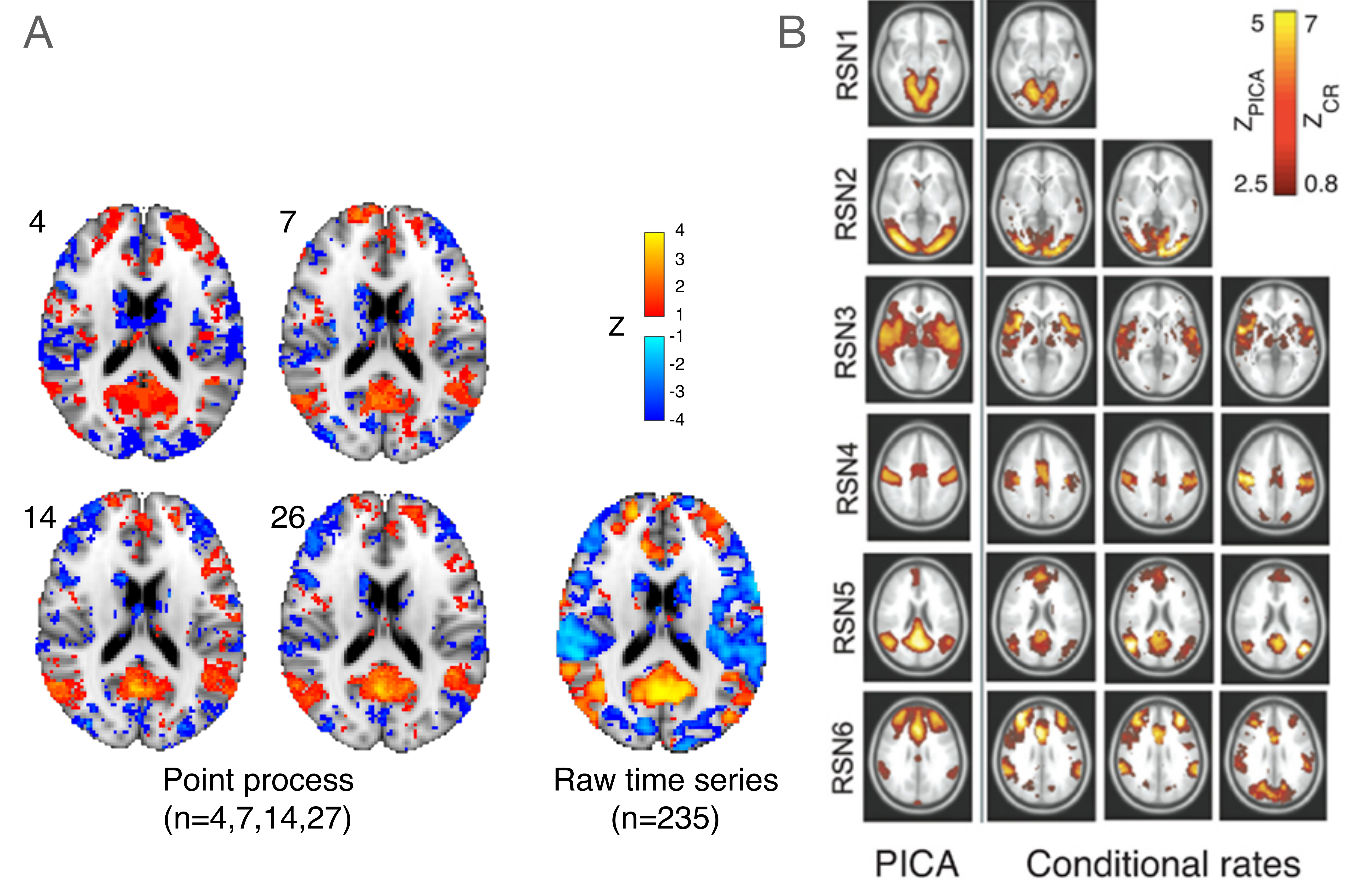} 
\caption{\footnotesize{Two examples of how a handful of points suffices to recover the functional connectivity between a given voxel and the rest of the brain. Panel A shows correlation maps obtained from the raw BOLD time-series of length n=235 (right image) and from the derived point process (left four images) using different 4,7,14 or 26 points. The left images represent, as ``heat maps", the co-activation of the seed (located at MNI coordinates x=4, y=-60, z=18) with each voxel. Note that a few points already suffice to identify well-defined clusters that are 1-4 standard deviations away from chance co-activations. Red/blue colors label positive/negative point co-activations in the case of the left maps and positive/negative correlations in the case of the right map. The left column images on Panel B represent the resting state maps of six resting-state networks obtained using PICA and the rightmost three columns the rate of points conditional to activity at a given seed (each column corresponds to a different seed, for further details see \cite{Tagliazucchi2012}). Scales for PICA (ZPICA) and conditional rate (ZCR) calculations are depicted in the inset. 
Panel A is reproduced from Ref. \cite{Cifre2020}, Panel B from \cite{Tagliazucchi2012}}}
\end{figure}
%%%%%%%%%%%%%%%%%%%%%%%%%%%%%%%%%%%%%%%%%%%%%%%%%%%%%%%%%%%%%%%%%%%%%%%%%%

\section{A few points suffice to compute functional connectivity}
The work described in the previous section demonstrated that the correlation between a small number of source and target events represents fairly well the functional connectivity estimates obtained from the Pearson correlation of the entire BOLD time series. The results of subsequent work~\cite{Tagliazucchi2016,Petridou2013,Allan2015,Li2014,Liu2013b,Liu2013,Chen2015,Jiang2014,Amico2014,Wu2013,Keilholz} provided ample support to our discovery by confirming the functional relevance of such relatively large amplitude BOLD events under variations of our original proposal.

Despite the substantial data reduction entailed by the point process, the information content of the few remaining points is very high. Figure 5A represents a qualitative comparison between the seed correlation maps obtained using the standard Pearson correlation method and the results using a few (4,7, 14, and 26) points. 
Notice that the method successfully identifies well-defined clusters that are several standard deviations away from chance co-activations.
A more complete comparison can be found in Tagliazucchi et al. \cite{Tagliazucchi2012} who demonstrated how the point process can extract the spatial location of six well-known resting-state network (RSN) maps. These networks describe the major independent components of spontaneous brain activity, and as such, they can be used as a relevant benchmark. The point process results were compared with the maps computed from the full BOLD time-series using probabilistic independent component analysis – PICA; \cite{Beckmann2005}, a well-established method. The heatmaps were done by calculating for six RSNs the rate of points co-occurrence (up to 2 time units later in this case) between representative sites (i.e., so-called “seeds”) and all other brain voxels. These results are presented as maps in Figure 5B (see\cite{Tagliazucchi2012} for details of the computation). The seeds locations for each of the six RSN were selected according to previous work. Figure 5B shows a striking similarity between the conditional rate maps and the respective PICA maps (rightmost three columns and left column of Figure 5B respectively) despite using less than 6\% of the raw fMRI information. Indeed, an average of about 5 seed points is enough to obtain RSN maps that are highly correlated (95\% confidence) with those obtained using PICA of the full BOLD signals.

\subsection{Why a few points suffice}

Despite the demonstration that a point process can represent very well the functional connectivity between ROIs, the underlying reason for the approach's success was unclear. 
Why can very few events (i.e., the time and location of the largest BOLD peaks) predict the functional connectivity between two ROIs?
The answer to this puzzle ended up being very simple: on any given signal the peaks and troughs represent a change in the ongoing tendency of the process (i.e., zero derivative) while, in contrast, the periods in between indicate redundancy, since the sign of their derivatives is maintained. Indeed, roughly speaking, peaks and troughs are the only data needed to determine the (instantaneous) periodicity of a time series. For instance, to track the cycles of economic expansion and contraction (or of a given stock price), the most informative points are the peaks and troughs of values. The same intuition applies to the functional connectivity case; most of the evolution of the BOLD signal can be reconstructed by interpolating between the location of its peaks and troughs. 
\begin{figure}[ht!]
\centering
\includegraphics [width = .6 \linewidth] {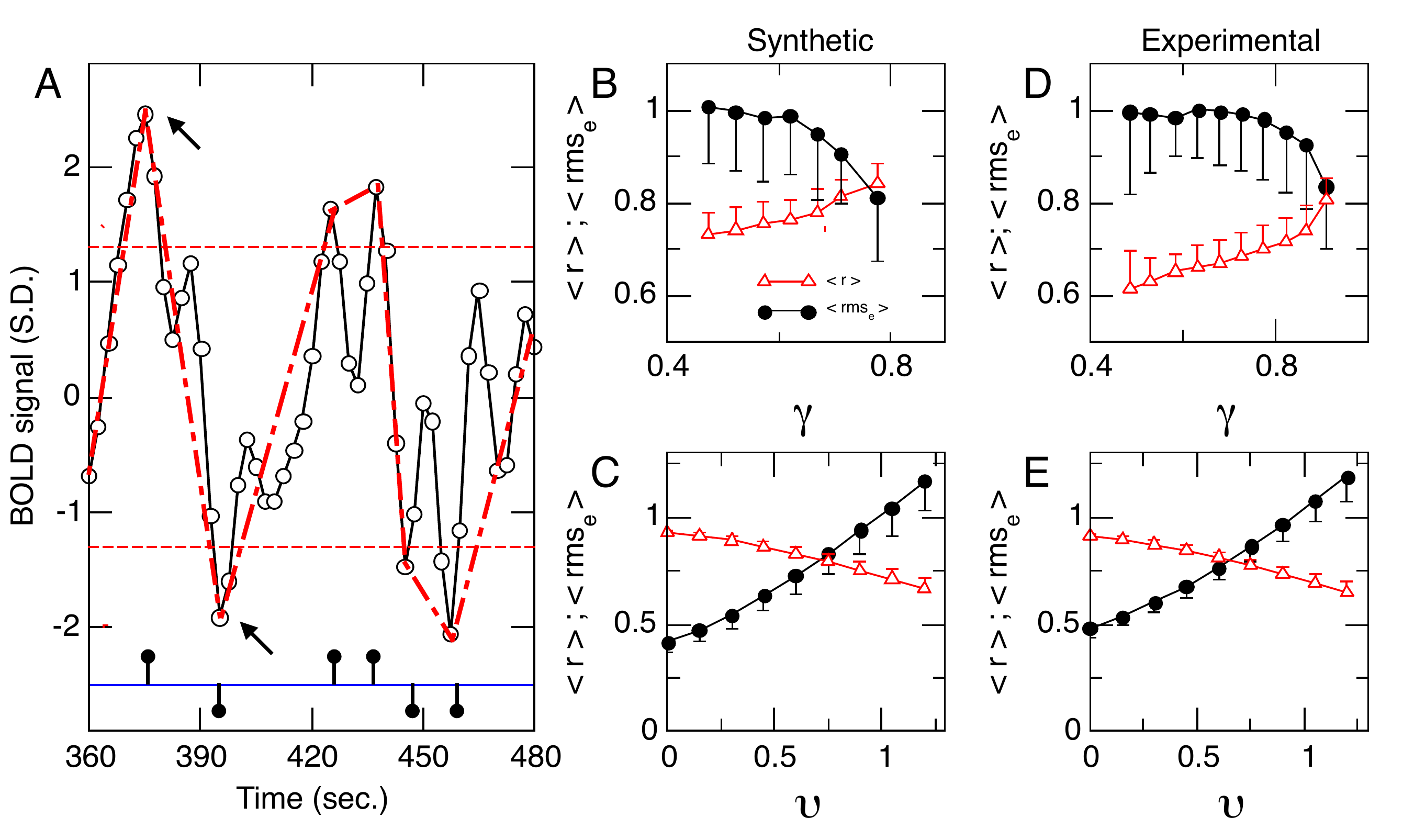} 
\caption{\footnotesize{The autocorrelation of the BOLD signal explains why the point process works: The trace in Panel A is an example of a resting-state recording of a BOLD brain signal. The point process is defined by the timing of the peaks and troughs larger than a given threshold (dotted lines; arrows indicate two of the peaks/troughs). Only six (depicted by the dots in the bottom trace) out of the 120 samples of the original time series are enough to preserve the same FC information as the entire original BOLD signal. This is due to its similarity with the piecewise linear time series (dashed red lines) defined by linear interpolation between the peaks and troughs. Panels D and E correspond to the computed similarity between the BOLD signals and the piecewise linear signals (evaluated by the correlation $\langle r \rangle$ and rmse values) for different autocorrelation $\gamma$ and threshold $\nu$ values. Panels B and C correspond to similar calculations using synthetic time series of various degrees of autocorrelation. For Panels B and D, the value of $\nu$ was fixed at 1. For Panels C and E, the value of $\gamma$ was 0.85}. Figure reproduced from Cifre el al. \cite{Cifre2020}.}
\end{figure}
%%%%%%%%%%%%%%%%%%%%%%%%%%%%%%%%%%%%%%%%%%%%%%%%%%%%%%%%%%%%%%%%%%%%%%%%%%

This argument was further elaborated recently by Cifre et al. \cite{Cifre2020} demonstrating that the key for the method lies in the temporal correlation properties of the time series under consideration. It was shown that signals with slowly decaying autocorrelations are particularly suitable for the point process to work. The results in Figure 6 summarize this analysis, in which the peaks and troughs of a BOLD time series define a piecewise linear approximation (Figure 6A) of the raw data. After that, the two signals (the raw and its piecewise approximation) were compared as a function of the threshold to define peaks and troughs. In Figs. 6D and 6E, the results are shown for different values of the threshold $\nu$ (in units of $\sigma$) and autocorrelation of the time-series, estimated by the value of the first autocorrelation coefficient $\gamma$. Panel D shows that when the BOLD signal's autocorrelation increases, the similarity between the piecewise linear and the raw signals increases as well. The similarity between the raw data and its linear piecewise approximation was evaluated in two ways: by the root mean squared error (RMSE) and by the linear correlation coefficient <r> between the two time series. As expected, raising the threshold $\nu$ above zero produces an increasing loss of information about the signal, which is reflected in a monotonic increase in the RMSE and a decrease in the <r> values (see Panel E).

\begin{figure}[ht!]
\centering
\includegraphics [width = .8 \linewidth] {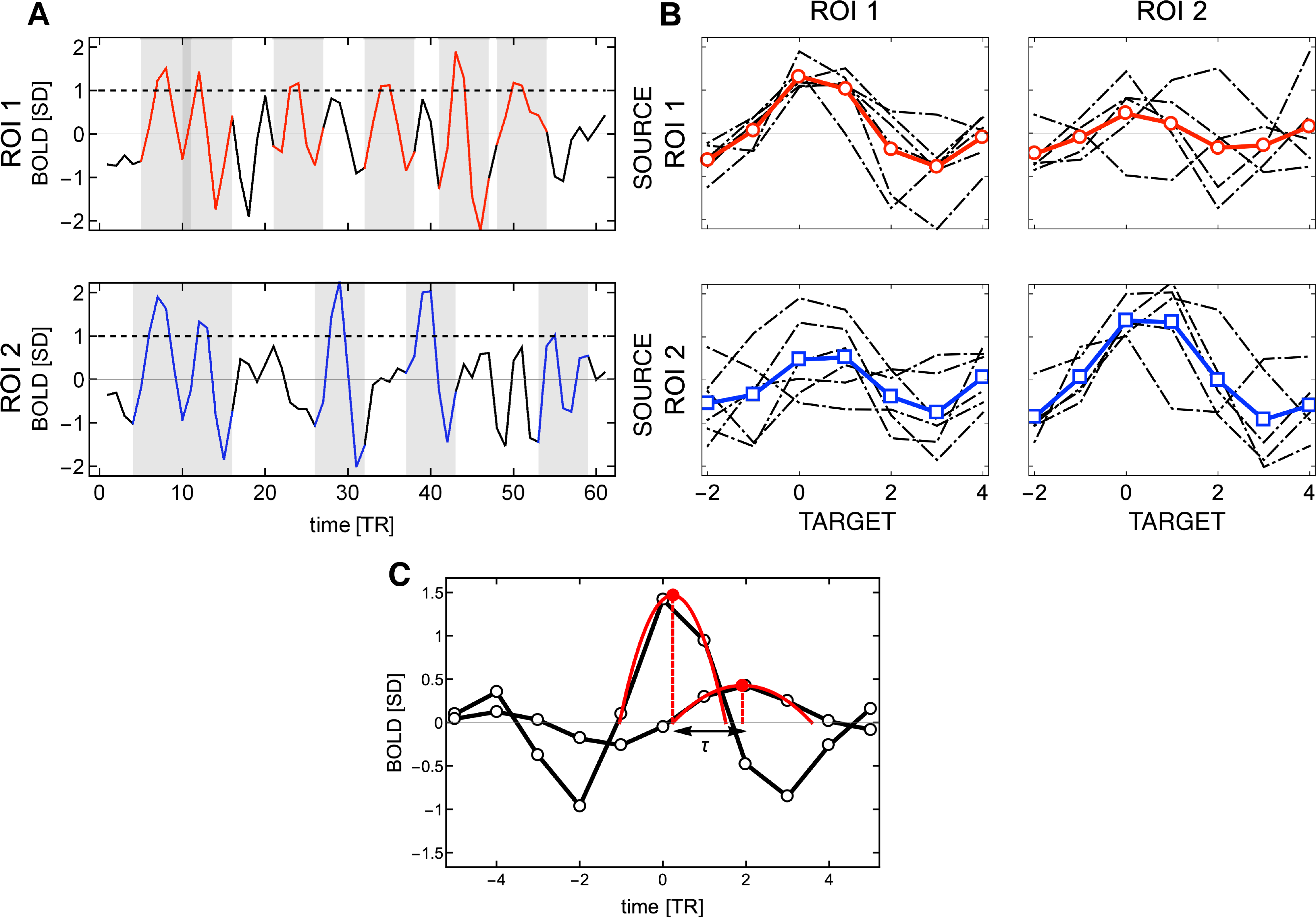} 
\caption{Definition and examples of events' directionality and delay. Panel A: The shaded areas highlight the timing of the source events (i.e., the BOLD signals surpassing the threshold). Notice that the source events of ROI 2 appear at different times than the source events of ROI 1 (e.g., around $TR=30$ and $TR=40$). Panel B: Shown are the individual events (in dot-dashed lines) and their averages (in red for ROI 1 and blue for ROI 2). The source events are shown in the top-left and bottom-right subplots, and the target events in the off-diagonal ones. Different sets of source events for each ROI give rise to asymmetry in the correlations between any two regions. Panel C: Cartoon representation of the delay $\tau$ estimation between two events with finer resolution than the original sampling TR values. First, the peak of the source event is centered at time $TR=0$ to estimate the closest peak of the target signal (here around time $TR=2$). After that, two parabolas are fitted to three sampled points in each of the peaks. The delay $\tau$ is defined by the time between the peaks of the two parabolas. Figure redrawn from \cite{Cifre2021}.}
 \label{ComputationsExplained}
\end{figure}

\section{Back to the future: the rBeta approach revisited}
In recent work \cite{Cifre2021} we revisited the rBeta proposal to describe unexplored features of the approach. In particular, we comment on the directed character of the rBeta events, which allows for a straightforward computation of the degree of asymmetry (i.e., directionality) of the correlation between two regions (Figure \ref{ComputationsExplained} Panels A and B). Since an event generated by a source ROI($i$) does not always imply an event by a target ROI($j$), the computation results in different $r(i,j)$ and $r(j,i)$ values when computing the event's correlation. In that way, the obtained matrix is non-symmetric. A quick measure of the asymmetry degree can be obtained from the average absolute difference of elements in the $r(i,j)$ matrix and its transpose.
Since by construction the method stores the timing for each event, it is possible to compute the delay between the timing of the source ROI to the closer target ROI as shown in the example presented in Figure \ref{ComputationsExplained}C. The information about directionality and delay accessible through rBeta may help expand the perspective of the usual functional connectivity paradigms into the realms of nonlinear time-dependent directed correlations.

\section{Final remarks}

To summarize, we have presented a brief chronological review of techniques to reduce fMRI brain data designed to facilitate interpretation of the functional connectivity. First, in section 2, we commented on how functional brain networks had been constructed from voxel-to-voxel correlations. Then, in Section 3 and 4, we presented the principles behind focusing solely on the large events by either extracting the rBeta events or converting the entire BOLD data set on a very compressed point process. The discussion included the reason for the success in using the large amplitude events, which was related to the autocorrelation of the BOLD signal. Finally, we commented briefly on some unexplored features offered by the rBeta technique, including the computation of directionality and delay, aspects that may deserve some attention in the future.

Overall, the common theme of our work has been to obtain a compressed and reduced data representation, able to help in the interpretation of the brain dynamics. In perspective, it may be worth contrasting this emphasis with the most recent approach, which focuses explicitly on the temporal evolution of the network edges \cite{Esfahlani,Sporns2021}.

This work recognizes (even in the title) our decade-old assertion that the functional connectivity is driven by the large amplitude events: \begin{quote}
\emph{Indeed, comparable results have been reported using similar methods \cite{Liu2013b,Tagliazucchi2012,Allan2015,Cifre2020, Zhang2020, Thompson2016, Betzel2016, Cabral2017, Thompson2015}. The principal finding of these studies is that high-amplitude activity is somehow related to stronger FC or the expression of particular brain systems.} \end{quote}

Sporns and colleagues start by building a time-series of the instantaneous edges {\emph before} the estimation of any given statistics. Any given time series corresponds to the instantaneous covariance between the activity of a pair of brain sites. After this initial step, the time series are binarized when the edges time series surpass given amplitude thresholds. The further analysis uses this binarized time series. As the authors remark in their publications, the end product of their analysis can not be very different from those extracted using our rBeta/point process methods. The important difference, however, is the very dissimilar computational burden of the two approaches. While in our approach {\emph before} any calculations, the entire data is reduced to a very small ($\sim$5-10\%) subset of events, the Sporns' approach \cite{Sporns2021} initially converts the original $N$ BOLD time-series into N*N edge's time-series. This initial {\emph expansion} of the data results in huge matrices of $\sim 10^12$ entries (i.e., $ 10^5 \times 10^5 voxels \times 10^2 TR $), a size that makes any voxel-wise exploration computationally prohibitive. Given this, it may be worth pursuing a careful comparison to determine the pros and cons of the two implementations.

%Final paragraph 
 % to be witen on a second pass
  
\section{Acknowledgments}  
Work conducted under the auspice of the Jagiellonian University-UNSAM Cooperation Agreement.
Financial  support was provided by the MICINN (Spain) grant PSI2017-82397-R and by CONICET (Argentina) and Escuela de Ciencia y Tecnolog\'ia, UNSAM.  JKO is supported in part by the Foundation for Polish Science (FNP) project ``Bio-inspired Artificial Neural Networks'' (POIR.04.04.00-00-14DE/18-00).

\bibliographystyle{unsrtnat}

\end{document}